\documentclass[
twocolumn,
letterpaper,
aps,
prl,
superscriptaddress,
floatfix,
longbibliography
]{revtex4-2}
\usepackage{graphicx}   
\usepackage{float} 
\usepackage{xspace}     
\usepackage{url}
\usepackage[mathlines]{lineno}
\usepackage{bm}
\usepackage{comment}
\usepackage{amsmath}
\usepackage{xcolor}
\usepackage[driverfallback=dvipdfm]{hyperref}
\usepackage{subfigure}
\hypersetup{
	colorlinks = true,
	citecolor= blue,
	linkbordercolor = cyan
}

\def \be {\begin{equation}}
\def \ee {\end{equation}} 
\def \bea {\begin{eqnarray}}
\def \eea {\end{eqnarray}}

\usepackage{color}
\definecolor{orange}{cmyk}{0.,0.353,1.,0.}    

\newcommand{\blue}[1]{{\color{blue}{#1}}}

\newcommand{\red}[1]{{\color{red}{#1}}}

\newcommand{\note}[1]{\textbf{\textcolor{magenta}{[#1]}}}
\def \new {\blue}

\usepackage{lineno}
\newcommand*\patchAmsMathEnvironmentForLineno[1]{
  \expandafter\let\csname old#1\expandafter\endcsname\csname #1\endcsname
  \expandafter\let\csname oldend#1\expandafter\endcsname\csname end#1\endcsname
  \renewenvironment{#1}
     {\linenomath\csname old#1\endcsname}
     {\csname oldend#1\endcsname\endlinenomath}}
\newcommand*\patchBothAmsMathEnvironmentsForLineno[1]{
  \patchAmsMathEnvironmentForLineno{#1}
  \patchAmsMathEnvironmentForLineno{#1*}}
\AtBeginDocument{
\patchBothAmsMathEnvironmentsForLineno{equation}
\patchBothAmsMathEnvironmentsForLineno{align}
\patchBothAmsMathEnvironmentsForLineno{flalign}
\patchBothAmsMathEnvironmentsForLineno{alignat}
\patchBothAmsMathEnvironmentsForLineno{gather}
\patchBothAmsMathEnvironmentsForLineno{multline}
}

\begin{document}

\preprint{\sl \new{version 8.0} \today. Text: \new{new}, \red{old}, \note{note}}

\title{Global polarization of $\Lambda$, $\Xi$, and $\Omega$ hyperons in Au+Au collisions from the RHIC Beam Energy Scan-II}

\author{The STAR Collaboration}


\begin{abstract} 


Global polarization of $\Lambda$, $\Xi$, and $\Omega$ hyperons is measured in Au+Au collisions at $\sqrt{s_{\mathrm{NN}}}=7.7$, 9.2, 11.5, 14.6, 17.3, 19.6, and 27~GeV using high-statistics data from the second phase of the RHIC Beam Energy Scan. 
No significant difference between the $\Lambda$ and $\bar{\Lambda}$ polarizations is observed over the measured collision-energy range, providing improved constraints on magnetic-field-induced contributions. 
The $\Lambda$ polarization increases with the magnitude of rapidity, with the slope of the linear fit differing from zero by 4.8 standard deviations, providing the first evidence for a nonzero rapidity dependence of global polarization in this energy range.
Global polarization of $\Xi^{-}+\bar{\Xi}^{+}$ is observed for the first time at these energies at a significance exceeding 5 standard deviations, with a magnitude comparable to that of $\Lambda+\bar{\Lambda}$ and an increase toward lower collision energies.
Measurements of $\Omega^{-}+\bar{\Omega}^{+}$ are compatible with a similar energy dependence and favor the decay parameter $\gamma_\Omega = +1$ with high probability under the assumption of the global polarization picture.
These measurements provide new constraints on the energy evolution, rapidity dependence, and species dependence of global polarization in heavy-ion collisions.

\end{abstract}

\maketitle




Quantum chromodynamics (QCD) predicts that hadronic matter undergoes a phase transition at extremely high temperatures and densities, forming a new state of matter called the quark-gluon plasma (QGP)~\cite{STAR:2005gfr, PHENIX:2004vcz,BRAHMS:2004adc,PHOBOS:2004zne,ALICE:2022wpn}.
Relativistic heavy-ion collisions at RHIC and the LHC provide a unique opportunity to study the properties of the QGP, which is found to be a near-perfect fluid~\cite{Heinz:2013th, Bernhard:2019bmu}. 
In non-central heavy-ion collisions, the large orbital angular momentum can induce strong vorticity in the produced medium, leading to hadron spin polarization along the direction of the system's angular momentum, known as global polarization~\cite{1,Gao:2007bc,2}.
Measurements of $\Lambda$ global polarization over collision energies from a few~GeV to several TeV have established global polarization as a sensitive probe of the vortical properties of the matter created in heavy-ion collisions~\cite{STAR:2007ccu,STAR:2017ckg,STAR:2018gyt,ALICE:2019onw,STAR:2021beb,HADES:2022enx,STAR:2023nvo,STAR:2022fan,STAR:2025dgs,Chen:2026gka}.

Several theoretical approaches based on thermal vorticity, spin--orbit correlations, and dynamical transport or hydrodynamic evolution have been developed to describe the observed global polarization and its collision-energy dependence~\cite{Becattini:2024uha,Karpenko:2016jyx, Vitiuk:2019rfv,Voronyuk:2023vyu,Yi:2024kwu,Sheng:2024pbw, Fang:2023bbw,Becattini:2020ngo,Niida:2024ntm,Chen:2024aom,Wu:2022mkr}.
Although these approaches reproduce important features of the existing $\Lambda$ measurements, the microscopic origin of global polarization and its dependence on collision energy, rapidity, and hadron species remain incompletely understood.

An important question is whether the strong magnetic field generated in relativistic heavy-ion collisions manifests itself through a splitting between the global polarizations of $\Lambda$ and $\bar{\Lambda}$. 
Such a splitting is expected due to their opposite magnetic moments and is sensitive to the magnitude and lifetime of the magnetic field at the time of hyperon formation (freeze-out)~\cite{Guo:2019joy,Becattini:2016gvu,Voronyuk:2011jd,Skokov:2009qp,Kharzeev:2007jp}. 
While previous measurements at $\sqrt{s_{\mathrm{NN}}}=19.6$ and 27~GeV~\cite{STAR:2023nvo} showed that polarizations of $\Lambda$ and $\bar{\Lambda}$ are statistically consistent, high-precision data at lower energies ($\sqrt{s_{\mathrm{NN}}}=7.7$--17.3~GeV) provide a unique opportunity to investigate possible polarization splitting further in a regime where magnetic-field effects are expected to be more pronounced due to the longer interaction time and the additional magnetic field potentially generated by the rotation of the positive net charge associated with baryon stopping.

The rapidity dependence of hyperon global polarization provides an additional test of its underlying dynamics~\cite{Baznat:2016rapidity,Sun:2017xhx,Liang:2019pst}. 
Hydrodynamic and transport models generally predict increasing polarization toward forward rapidities~\cite{Wu:2019eyi,Jiang:2016woz,Deng:2016gyh,Liang:2019pst}, whereas other approaches predict a decreasing trend at larger rapidities~\cite{Guo:2021udq,Wei:2018zfb,Xie:2019jun,Liang:2019pst}. 
Previous measurements largely restricted to $|\eta|<1.0$ have not yet provided conclusive constraints on the rapidity dependence because of limited acceptance and statistical precision~\cite{STAR:2017ckg,STAR:2018gyt,ALICE:2019onw,STAR:2021beb,HADES:2022enx,STAR:2023nvo}. 
The inner Time Projection Chamber (iTPC) upgrade extended the STAR tracking acceptance to $|\eta|<1.5$~\cite{Shen:2018pkc}, and BES-II provides approximately 20 times the statistics of BES-I. 
These measurements therefore provide improved constraints on the rapidity dependence of global polarization.

The species dependence of global polarization provides a complementary test of the polarization mechanism, in particular whether the behavior observed for $\Lambda(uds)$ extends to multistrange hyperons such as $\Xi(dss)$ and $\Omega(sss)$. 
The multiphase transport model (AMPT) calculations predict $P_{\Omega}>P_{\Xi}>P_{\Lambda}$ considering the spin $3/2$ nature of $\Omega$ and decay feed-down~\cite{Li:2021zwq}. 
Independent calculations incorporating quark--quark spin correlations show that these correlations can modify the relation $P_{\Omega}>(5/3)P_{\Lambda}$~\cite{Zhang:2024hyq,Oliva:2026wbo}. 
Previous measurements of $\Xi$ and $\Omega$ polarization are available only for $\sqrt{s_{\mathrm{NN}}}=200~\mathrm{GeV}$ with limited statistics~\cite{STAR:2020xbm, STAR:2025dgs}, leaving their collision-energy dependence poorly constrained experimentally. 
The high-precision measurements of $\Lambda$, $\Xi$, and $\Omega$ polarization using the BES-II data provide an opportunity to test the above predictions and whether a common vorticity-induced spin--orbit-coupling mechanism describes the polarization of different hyperons.



In this paper, we present measurements of the global polarization of $\Lambda$, $\bar{\Lambda}$, $\Xi^{-}$/$\bar{\Xi}^{+}$, $\Omega^{-}$/$\bar{\Omega}^{+}$ hyperons in Au+Au collisions at $\sqrt{s_{\mathrm{NN}}}=7.7$, 9.2, 11.5, 14.6, 17.3, 19.6, and 27~GeV. 
The measurements are reported as a function of collision energy. Additionally, the $\Lambda$ polarization at 20--50~\% centrality and for transverse momentum $0.5 < p_T < 6.0$~GeV/c is reported as a function of rapidity.
  
The data were collected by the STAR experiment~\cite{STAR:2002eio} during Phase II of the Beam Energy Scan program (BES-II) with the upgraded detector at RHIC.
The collision vertices were reconstructed using charged-particle tracks measured in the TPC~\cite{Anderson:2003ur,Shen:2018pkc}.
Events were selected based on the collision vertex position along the beam direction within (-145, 145)~cm for $\sqrt{s_{\mathrm{NN}}}=7.7$--19.6~GeV and within (-70, 70)~cm for $\sqrt{s_{\mathrm{NN}}} = 27$~GeV, and within $\pm2$~cm in the radial direction, relative to the beam center. 
The out-of-time pile-up events, a beam-induced background where multiple collisions are recorded as one event, were suppressed by excluding outliers in the correlation between the total number of TPC tracks and the number matched with a hit in the Time Of Flight (TOF) detector~\cite{Llope:2012zz}.

These selection criteria yielded about 110, 204, 269, 413, 275, 668, and 521 million minimum-bias (MB) events for those datasets. 
The MB trigger requires hits of both vertex position detectors
(VPD)~\cite{Llope:2014nva} located at forward and backward pseudorapidities ($4.23<|\eta|<5.1$) and in the zero-degree calorimeters (ZDCs)~\cite{Adler:2001fq}, which detect spectator neutrons at $|\eta|>6.3$.
The statistics at 7.7, 11.5, and 14.6~GeV are more than 10 times higher than those of the previous BES-I datasets~\cite{STAR:2017ckg}, 
with two new energies at 9.2 and 17.3~GeV also included.
The centrality classes are determined using the charged-particle multiplicity measured in the TPC within the pseudorapidity range $|\eta|<0.5$ and are defined through comparison with Monte Carlo Glauber model simulations~\cite{STAR:2012och}.

The reconstruction and selection procedures for $\Lambda$($\bar{\Lambda}$), $\Xi^{-}$($\bar{\Xi}^{+}$) and $\Omega^{-}$($\bar{\Omega}^{+}$) hyperon candidates are based on the procedures described in Refs.~\cite{STAR:2017ckg,STAR:2023nvo,STAR:2020xbm}.
The charged-particle tracks are reconstructed with the TPC. The ionization energy loss in the TPC gas and the flight-time information from the TOF are used to identify charged-particle track species ($\pi$, $K$, and $p$).
The pseudorapidity coverage of TPC tracks is $|\eta| < 1.0$ for the 27~GeV dataset. With the upgrade of the inner TPC readout, the accepted pseudorapidity range was extended from $|\eta| < 1.0$ to $|\eta| < 1.5$ for the 7.7--19.6~GeV datasets. Concurrently, the lower cutoff for reconstructed track transverse momentum was reduced from $p_T > 0.2$~GeV/$c$ to $p_T > 0.15$~GeV/$c$ for these datasets~\cite{Shen:2018pkc}.
Then the $\Lambda$($\bar{\Lambda}$), $\Xi^{-}$($\bar{\Xi}^{+}$), and $\Omega^{-}$($\bar{\Omega}^{+}$) candidates were reconstructed utilizing the decay channels $\Lambda \to p+\pi^{-}$, $\Xi^{-}\to \Lambda\pi^{-}$, and $\Omega^{-}\rightarrow \Lambda K^{-}$.
The invariant mass was calculated for all $p$-$\pi$, $\Lambda$-$\pi$, and $\Lambda$-$K$ pairs to find $\Lambda$, $\Xi$, and $\Omega$ candidates, respectively.  
Details on the reconstruction can be found in Ref.~\cite{longpaper}.


The global hyperon polarization $P_{\rm H}$, along the initial angular momentum direction and relative to the reaction plane, can be determined as follows~\cite{STAR:2007ccu}:
\begin{equation}
    P_{\rm H} = \frac{8}{\pi\alpha_{\rm H}} \frac{1}{A_0} \frac{\langle\sin(\Psi_1-\phi_{\rm D}^{\ast})\rangle}{\rm Res(\Psi_1)}.
    \label{eq:PH}
\end{equation}
Here, $\alpha_{\rm H}$ is the hyperon decay parameter, with $\alpha_{\Lambda}=-\alpha_{\bar{\Lambda}}=0.732\pm0.014$, $\alpha_{\Xi^{-}}=-\alpha_{\bar{\Xi}^{+}}=-0.401\pm0.010$, and $\alpha_\Omega = 0.0157 \pm 0.0021$~\cite{ParticleDataGroup:2020ssz}. The acceptance correction factor $A_0$ is estimated from the data and depends on centrality, $p_{\rm T}$, and $\eta$ because of the nonuniform acceptance of the STAR detector~\cite{STAR:2007ccu}. $\phi_{\rm D}^{\ast}$ is the azimuthal angle of the daughter baryon in the hyperon rest frame. $\Psi_1$ denotes the first-order event plane angle, which provides an experimental estimate of the reaction plane orientation, and ${\rm Res}(\Psi_1)$ denotes the corresponding event plane resolution.

The reaction plane is defined by the beam axis and the impact-parameter vector connecting the centers of the two colliding nuclei, with the system's orbital angular momentum oriented perpendicular to it~\cite{Adams:2021yob}. Experimentally, $\Psi_1$ is reconstructed from the azimuthal distribution of spectator nucleons and forward-going particles measured by the Event Plane Detector (EPD)~\cite{Adams:2019fpo}, which covers the forward and backward pseudorapidity regions $2.1<|\eta|<5.1$. The event plane resolution ${\rm Res}(\Psi_1)$ quantifies how well $\Psi_1$ estimates the reaction plane orientation and is determined from the correlation between the $\Psi_1$ measurements in the west and east EPD at forward and backward rapidities~\cite{Poskanzer:1998yz,Voloshin:2008dg}. 
The EPD provides a larger ${\rm Res}(\Psi_1)$ than the Beam Beam Counter (BBC) used in the BES-I measurements~\cite{STAR:2017ckg}, thereby reducing the uncertainty of the measured global polarization.


The extraction of $\langle\sin(\Psi_1-\phi_{\rm D}^{\ast})\rangle$, namely the direct measurement of $\Lambda$ and $\Xi$ polarization, was performed in the same way as in previous studies~\cite{STAR:2017ckg, STAR:2021beb, STAR:2023nvo,STAR:2020xbm}.
For $\Xi$ and $\Omega$, their polarization can also be measured indirectly via the daughter-$\Lambda$ polarization through the spin transfer factor in the decay process ($\Xi^{-}\rightarrow \Lambda\pi^{-}$, $\Omega^{-}\rightarrow \Lambda K^{-}$), $P_{\Xi(\Omega)} = P_{\Lambda}/C_{\Xi\Lambda(\Omega\Lambda)}$~\cite{STAR:2020xbm, longpaper}.
For the $\Xi^{-}\rightarrow \Lambda\pi^{-}$ process,
$C_{\Xi\Lambda}=\frac{1}{3}(1+2\gamma_\Xi)=0.944$, with the measured weak-decay parameter $\gamma_\Xi=0.916$~\cite{E756:2000rge,HyperCP:2004not,ParticleDataGroup:2020ssz, STAR:2020xbm}.
The polarization of the daughter baryon in a weak decay of a spin $J=3/2$ hyperon like $\Omega$ can be described by three decay parameters, $\alpha_\Omega$, $\beta_\Omega$, $\gamma_\Omega$, satisfying $\alpha^2_\Omega + \beta^2_\Omega + \gamma^2_\Omega = 1$. 
The time-reversal violation parameter $\beta_\Omega$ is expected to be small, which limits the unknown decay parameter $\gamma_\Omega$ to $\pm1$. 
This results in a polarization transfer coefficient of either $C_{\Omega\Lambda} \approx 1.0$ or $C_{\Omega\Lambda} \approx -0.6$, where $C_{\Omega\Lambda}=\frac{1}{5} [1+4\gamma_\Omega]$~\cite{STAR:2020xbm}. 
Given the tiny value of $\alpha_\Omega = 0.0157 \pm 0.0021$, a direct measurement of the $\Omega$ polarization via the angular distribution of the daughter $\Lambda$ is practically impossible. 
Thus $\Omega$ polarization is measured through the spin transfer factor $C_{\Omega\Lambda}$ in the decay process.

The sources of systematic uncertainty in the polarization measurement include the polarization contribution from background hyperon candidates, variation of the hyperon reconstruction selection criteria, event plane determination, event plane resolution, acceptance effect, and the uncertainty on the hyperon decay parameter.
Systematic effects related to the polarization contribution of background hyperon candidates are investigated by changing the background probability density function from a second-order to a first-order polynomial function.
The resulting background systematic uncertainties are approximately 2~\% for $\Lambda$, 8~\% for $\Xi$, and 30~\% for $\Omega$. 
The larger uncertainty for $\Omega$ arises from its lower signal-to-background ratio and the subtraction of residual polarization from the remaining background, with detailed source-by-source decompositions provided in Ref.~\cite{longpaper}.

The systematic uncertainties from the hyperon reconstruction selection criteria are evaluated by varying the topological cuts in the reconstruction of $\Xi$ and $\Omega$, yielding uncertainties of $\sim$3~\% for the direct $\Xi$ measurement, $\sim$4~\% for the indirect $\Xi$ measurement, and $\sim$6~\% for $\Omega$.
The systematic uncertainties from event-plane determination are estimated by comparing $\Psi_1$ reconstructed from a single EPD side (East or West) with that from the full EPD (both sides).
The relative $P_H$ differences are evaluated for each centrality bin and averaged within the 0--20~\%, 20--50~\%, and 50--80~\% centrality intervals separately at each collision energy. The resulting uncertainty is applied to all measurements in the corresponding interval at that energy. Averaging these uncertainties over the BES-II energies yields representative values of approximately 15.6~\%, 3.5~\%, and 4.0~\%, respectively.

The uncertainty in the hyperon decay parameter $\alpha_{\rm H}$ results in a relative systematic uncertainty of $\sim$1.9~\% for the $\Lambda$ and the indirect $\Xi$ and $\Omega$ measurements,
and $\sim$2.5~\% for the direct $\Xi$ measurement~\cite{ParticleDataGroup:2020ssz}.
The systematic uncertainties from the event plane resolution and acceptance correction $A_{0}$ are each less than 0.1~\%.
The total systematic uncertainties were calculated by adding the contributions from each source in quadrature.
We note that the polarization measurements are not corrected for feed-down contributions from particle decays.

For comparisons among hyperon species, the systematic uncertainties associated with event-plane determination and resolution are treated as correlated and propagated accordingly when evaluating polarization differences. The uncertainties associated with acceptance corrections, decay parameters, and background subtraction are evaluated separately for each hyperon species and measurement method.

\begin{figure}[htb]
    \centering
    \includegraphics[width=0.95\linewidth]{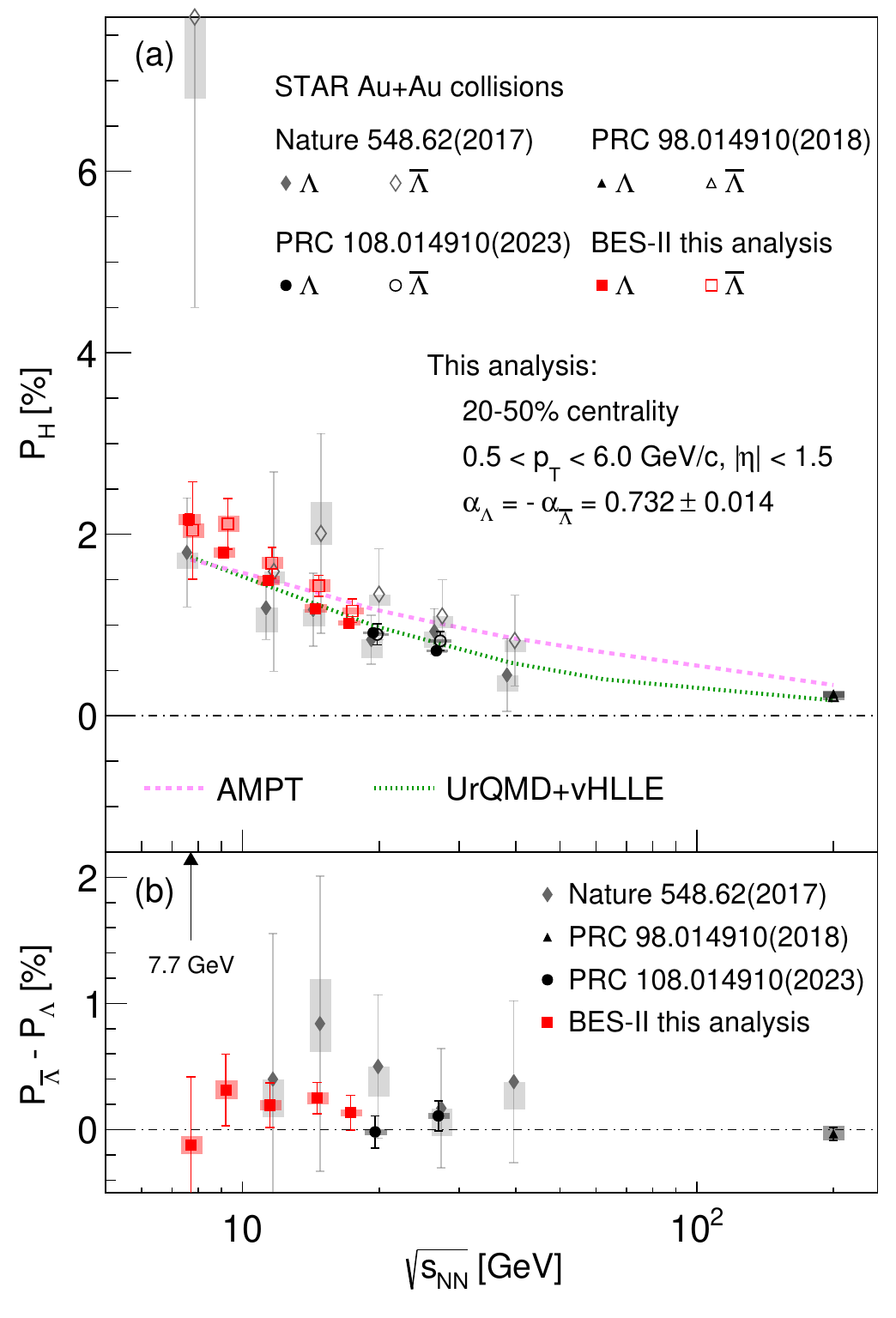}
    \caption{The global polarization of $\Lambda$ and $\bar{\Lambda}$ in Au+Au collisions as a function of collision energy at $\sqrt{s_{\mathrm{NN}}}=7.7$, 9.2, 11.5, 14.6, and 17.3~GeV from STAR BES-II. 
    The mid-central (20--50~\%) $P_{H}$ results are shown in panel (a), and the difference between $P_{\bar{\Lambda}}$ and $P_{\Lambda}$ is shown in panel (b), as a function of collision energy together with previous measurements. 
    Previous STAR results~\cite{STAR:2017ckg,STAR:2018gyt} compared here are rescaled using the new decay parameters, as shown in the figure.
    The polarization results are compared with two theoretical calculations that include feed-down effects~\cite{Karpenko:2016jyx, Li:2021zwq}.
    Open boxes and vertical lines represent systematic and statistical uncertainties, respectively.}
    \label{fig:GlobalP_Lam_AntiLam}
\end{figure}

Figure~\ref{fig:GlobalP_Lam_AntiLam} shows the global polarization of inclusive $\Lambda$ and $\bar{\Lambda}$ hyperons as a function of collision energy integrated over 20--50~\% centrality in Au+Au collisions at $\sqrt{s_{\mathrm{NN}}}=7.7$, 9.2, 11.5, 14.6, and 17.3~GeV from STAR BES-II.
The results are also compared with previous BES-I measurements~\cite{STAR:2017ckg}, showing a substantial improvement in statistical precision, with the statistical uncertainties reduced by approximately a factor of 10.
The new BES-II results therefore provide a much clearer picture of the collision-energy dependence of the global polarization for both $\Lambda$ and $\bar{\Lambda}$.

The $\Lambda$ polarization data are compared with calculations based on two models: the viscous hydrodynamic model UrQMD+vHLLE~\cite{Karpenko:2016jyx} and the string-melting AMPT model~\cite{Li:2021zwq}.
Both models consider the feed-down contribution from heavier hyperon decays and can qualitatively describe the decreasing trend versus collision energy. 
The AMPT model attributes the decreasing behavior to smaller vorticity generated in the mid-rapidity region at higher collision energies, while UrQMD+vHLLE explains the enhanced polarization at lower collision energies with an emerging shear-flow pattern in the beam direction related to baryon stopping. 
As seen in the upper panel of Fig.~\ref{fig:GlobalP_Lam_AntiLam}, UrQMD+vHLLE results reproduce the data at $\sqrt{s_{\mathrm{NN}}}\geq11.5$~GeV, while AMPT results overshoot the data points at $\sqrt{s_{\mathrm{NN}}}\geq14.6~\mathrm{GeV}$.
We note that both the AMPT and UrQMD+vHLLE models underestimate the $\Lambda$ data at collision energies below 10~GeV, which indicates the need for further investigation of polarization mechanisms in the low energy region.

The lower panel of Fig.~\ref{fig:GlobalP_Lam_AntiLam} shows the difference between $\Lambda$ and $\bar{\Lambda}$, $\Delta P=P_{\bar{\Lambda}}-P_{\Lambda}$, as a function of collision energy. With significantly reduced uncertainties, the difference $\Delta P$ is further confirmed to be consistent with zero across a wider energy range. 
$\Delta P$ is related to the strength of the magnetic field due to the opposite magnetic moments of $\Lambda$ and $\bar{\Lambda}$, under a thermal approach~\cite{Becattini:2016gvu,Guo:2019joy, Muller:2018ibh}.
Consequently, the effective magnetic field at the time of $\Lambda$ emission is estimated as
$|B| \approx (T_s |P_{\bar{\Lambda}} - P_{\Lambda}|)/(2|\mu_\Lambda|)$~\cite{Becattini:2016gvu},
where $T_s=150$~MeV is the assumed emission temperature and
$\mu_\Lambda=-1.93\times10^{-14}$~MeV/T is the $\Lambda$ magnetic moment. 
Within the above thermal freeze-out assumptions, the extracted magnetic fields are consistent with zero, and the corresponding upper limits at the 95~\% confidence level are summarized in Table~\ref{MEuplimit}~\cite{STAR:2023nvo}.

\begin{table}[ht]
    \centering
    \caption{Upper limits on the magnetic field}
    \begin{tabular}{c|c|c|c|c|c|c|c}
        \toprule
        $\sqrt{s_{\mathrm{NN}}}$ (GeV)  & 7.7 & 9.2 & 11.5 & 14.6 & 17.3 & 19.6 & 27 \\
        \hline
        B ($\times 10^{13}$ T) & 3.8 & 3.5 & 2.2 & 2.0 & 1.6 & 0.9 & 1.4 \\
        \hline
        \hline
    \end{tabular}
    \label{MEuplimit}
\end{table}

\begin{figure}[htb]
    \centering
    \vspace{-2mm}
    \includegraphics[width=0.95\linewidth]{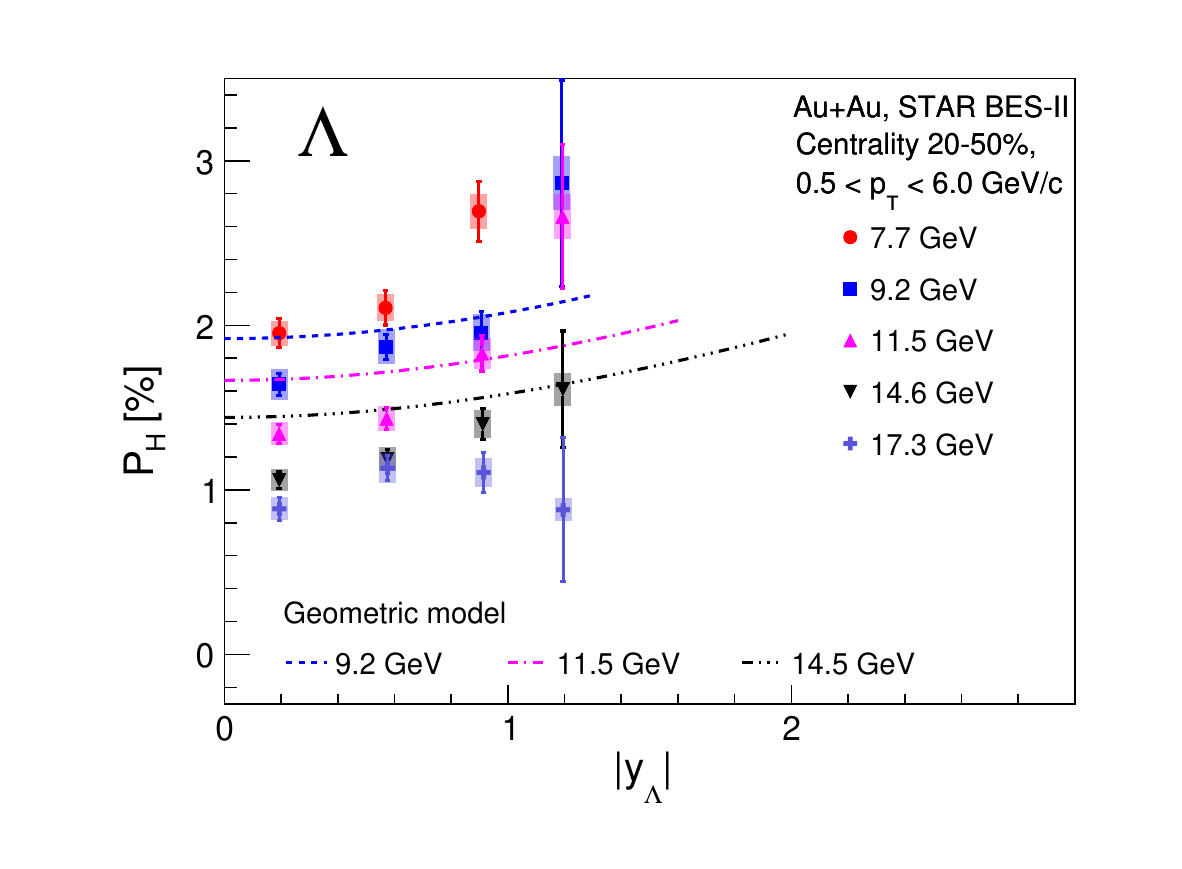}
    \vspace{-10mm}
    \hspace{3mm}
    \caption{Global polarization of $\Lambda$ as a function of $|y|$ for 20--50~\% centrality in Au+Au collisions at $\sqrt{s_{\mathrm{NN}}}=7.7$, 9.2, 11.5, 14.6, and 17.3~GeV. The results are also compared with the geometric model calculation, shown as curves~\cite{Liang:2019pst}.}
    \label{fig:GlobalP_Eta}
\end{figure}

The rapidity dependence of $\Lambda$ polarization is shown in Fig.~\ref{fig:GlobalP_Eta} as a function of the absolute value of the hyperon rapidity, $|y|$. 
Only the $\Lambda$ results are presented here because the $\bar{\Lambda}$ measurements have limited statistical precision at larger $|y|$.
The polarization increases with $|y|$, and the increasing trend becomes more pronounced toward lower collision energies. 
To quantify the rapidity dependence, the measurements at $\sqrt{s_{\mathrm{NN}}}=7.7$--17.3~GeV were first combined within each $|y|$ interval and then fitted with a linear function, $P=a+b|y|$. The fit yields a slope of $b=0.498\pm0.103$, including both statistical and systematic uncertainties, which differs from zero by $4.8$ standard deviations.

The data are compared with a geometric model calculation~\cite{Liang:2019pst},
as shown by the curves for three energies (9.2, 11.5, 14.6~GeV) in Fig.~\ref{fig:GlobalP_Eta}.
With rapidity-dependent transfer coefficients from the orbital angular momentum in the initial state to final-state hyperon polarization, the geometric model predicts the increasing trend of $\Lambda$ polarization versus rapidity, but with a weaker rapidity dependence than observed in the data~\cite{Liang:2019pst}.
In contrast, a chiral kinetic calculation using initial conditions from the AMPT model~\cite{Sun:2017xhx} predicts that, at $\sqrt{s_{\mathrm{NN}}}=7.7$~GeV, the polarization peaks at zero rapidity and decreases with increasing $|y|$, attributed to the smaller number of scatterings experienced by quarks and antiquarks at large rapidity.
The Particle-in-Cell Relativistic (PICR) hydrodynamics model
calculation~\cite{Xie:2019jun} at $\sqrt{s_{\mathrm{NN}}}=200$~GeV
predicts a polarization that has a flat rapidity dependence for
$|y|<1.0$ and drops rapidly to zero for $|y|>1.0$.
Our data on the rapidity dependence discriminate qualitatively among current calculations, and further investigation will help to fully understand the global polarization mechanism.
A more detailed study of the centrality, transverse-momentum, and rapidity dependences, including comparisons with calculations based on
the SMASH~IC+CLVisc framework~\cite{Li:2026udm}, is presented in Ref.~\cite{longpaper}.

\begin{figure}[htb]
    \centering
    \includegraphics[width=0.95\linewidth]{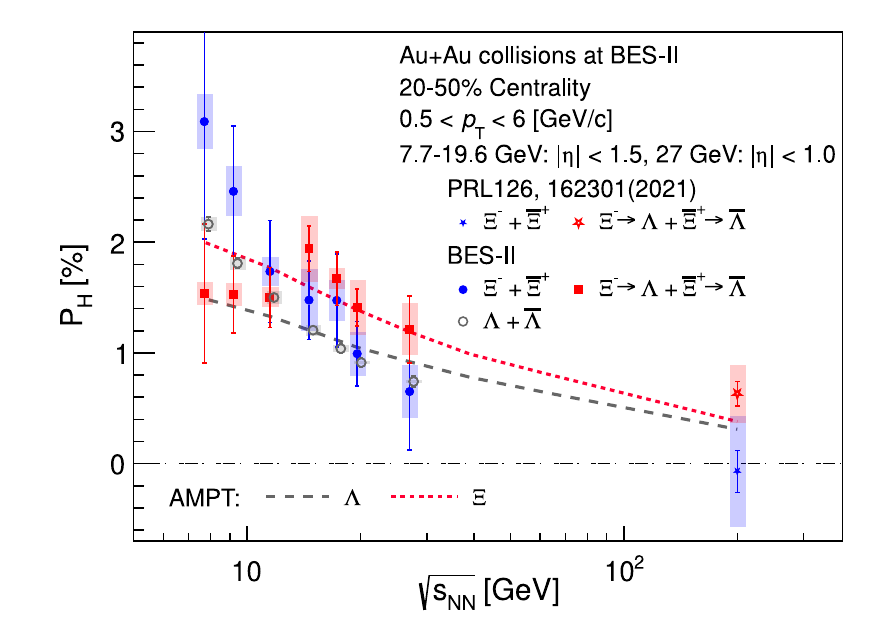}
    \hspace{3mm}
    \caption{ Global polarization of $\Xi^{-}+\bar{\Xi}^{+}$ as a function of collision energy in Au+Au collisions at $\sqrt{s_{\mathrm{NN}}}=7.7$, 9.2, 11.5, 14.6, 17.3, 19.6, and 27~GeV. Results for the inclusive $\Lambda$+$\bar{\Lambda}$ polarization are shown for comparison. The direct $\Xi$ measurement (blue markers) and the indirect measurement via the daughter-$\Lambda$ polarization (red markers) are both shown.}
    \label{fig:GlobalP_Xi}
\end{figure}
Figure~\ref{fig:GlobalP_Xi} presents the global polarization of $\Xi^{-}$+$\bar{\Xi}^{+}$ as a function of collision energy in Au+Au collisions at $\sqrt{s_{\mathrm{NN}}}=7.7$, 9.2, 11.5, 14.6, 17.3, 19.6, and 27~GeV from BES-II. 
Since no significant difference is observed between the polarization of $\Xi^{-}$ and $\bar{\Xi}^{+}$, the results are combined to reduce statistical uncertainty. 
A clear polarization signal of $\Xi^{-}+\bar{\Xi}^{+}$ is observed and exhibits a collision energy dependence, rising toward lower energies. 
Using inverse-statistical-variance weighting over $\sqrt{s_{\mathrm{NN}}}=7.7$--27~GeV and propagating correlated and uncorrelated systematic uncertainties separately, the energy-averaged polarizations are $P_{\Xi^{-}+\bar{\Xi}^{+}}=1.389\pm0.163~(\mathrm{stat.})\pm0.212~(\mathrm{syst.})~\%$ for the direct measurement and $P_{\Xi^{-}\rightarrow\Lambda+\bar{\Xi}^{+}\rightarrow\bar{\Lambda}}=1.565\pm0.094~(\mathrm{stat.})\pm0.200~(\mathrm{syst.})~\%$ for the indirect measurement, corresponding to $5.2\sigma$ and $7.1\sigma$, respectively.
The two measurements are consistent with each other, with a difference less than $1\sigma$. 
The inclusive $\Lambda$+$\bar{\Lambda}$ polarization is also shown for comparison.
At the current precision, the $\Xi^{-} + \bar{\Xi}^{+}$ polarization shows no significant difference relative to the $\Lambda + \bar{\Lambda}$ polarization.

The AMPT~\cite{Li:2021zwq} calculation predicts a decreasing $\Xi$
polarization with increasing collision energy and can qualitatively
describe the measurement results.
In this framework, hadronization proceeds through quark coalescence,
and the hyperon polarization is calculated from the local thermal
vorticity at hadronization, giving comparable primary polarizations,
$P_{\Xi}\simeq P_{\Lambda}$.
In an alternative quark-coalescence scenario, the quarks generated in
the QGP are assumed to be globally polarized due to spin--orbit coupling and then combine into hyperons, also giving comparable primary
polarizations if strange and light quarks have similar polarizations~\cite{Li:2021zwq}.
The AMPT calculation further includes feed-down effects, which increase
the $\Xi^{-}$ polarization by about 25~\% and reduce the $\Lambda$
polarization by about 10~\% relative to their primary
polarizations~\cite{Li:2021zwq}.
This results in the predicted hierarchy $P_{\Xi}>P_{\Lambda}$ shown in Fig.~\ref{fig:GlobalP_Xi}.
Although the experimental data are not corrected for feed-down effects, they are consistent with the AMPT calculation within uncertainties; however, the predicted difference between $\Xi$ and $\Lambda$ polarization is not resolved at the current experimental precision.

\begin{figure}[htb]
    \centering
    \includegraphics[width=0.95\linewidth]{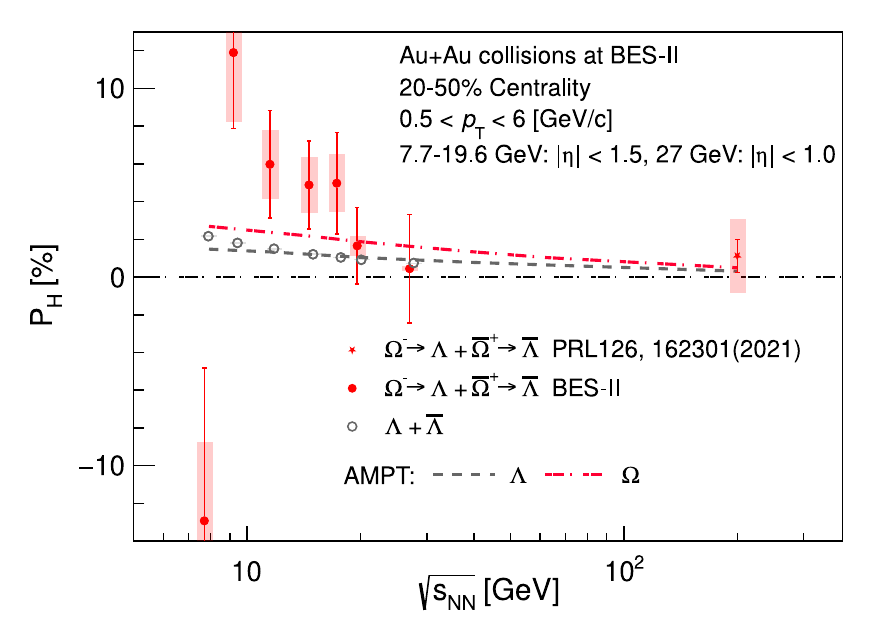}
    \hspace{3mm}
    \caption{Global polarization of $\Omega^{-}+\bar{\Omega}^{+}$ as a function of collision energy in Au+Au collisions at $\sqrt{s_{\mathrm{NN}}}=7.7$, 9.2, 11.5, 14.6, 17.3, 19.6, and 27~GeV. Results for the inclusive $\Lambda$+$\bar{\Lambda}$ polarization are shown for comparison. }
    \label{fig:GlobalP_Omega}
\end{figure}

Figure~\ref{fig:GlobalP_Omega} presents the global polarization of the $\Omega^{-}$+$\bar{\Omega}^{+}$ as a function of collision energy at 20--50~\% centrality in Au+Au collisions using BES-II data. 
The $\Omega^{-}$+$\bar{\Omega}^{+}$ polarization is measured via the daughter-$\Lambda$ polarization under the assumption of $C_{\Omega\Lambda} = 1$~\cite{STAR:2020xbm}. 
Despite the negative value at the lowest collision energy, the $\Omega^{-}+\bar{\Omega}^{+}$ polarization shows a decreasing trend with increasing collision energy.
The inclusive $\Lambda+\bar{\Lambda}$ polarization is also shown in Fig.~\ref{fig:GlobalP_Omega} for comparison. 
There is a hint of larger $\Omega^{-}+\bar{\Omega}^{+}$ polarization than that of $\Lambda+\bar{\Lambda}$ at lower collision energies. 
Integrated over the BES-II energy range (7.7--27~GeV), the average polarizations are $P_{\Omega^{-} + \bar{\Omega}^{+}}=3.731\pm1.062~(\mathrm{stat.})\pm1.289~(\mathrm{syst.})~\%$ and $P_{\Lambda + \bar{\Lambda}}=1.284\pm0.017~(\mathrm{stat.})\pm0.039~(\mathrm{syst.})~\%$.
The quantitative difference is about $1.5\sigma$ after considering all the uncertainties.

We note that the $\Omega$ polarization will flip sign with the spin transfer factor of $C_{\Omega\Lambda} \approx -0.6$ assuming $\gamma_\Omega \approx -1$, thus deviating dramatically from $\Lambda$ and $\Xi$ polarization.  
Assuming similar global polarization for all the quarks due to spin--orbit coupling, a factor of $5/3$ is expected for $\Omega$ polarization compared to $\Lambda$ polarization, as $\Omega$ has spin-$3/2$~\cite{Li:2021zwq}.
The quantitative comparison of measured $\Omega$ and inclusive $\Lambda$ polarization after considering the $5/3$ factor gives $\chi^2/\mathrm{NDF}=9.25/7$ ($p=0.24$) for $\gamma_\Omega \approx 1$ and
$22.81/7$ ($p=1.8\times10^{-3}$) for $\gamma_\Omega \approx -1$, respectively.
Under this assumption, the measurements favor the $\gamma_\Omega\approx+1$ solution.

The $\Omega$ data are also compared with AMPT calculations, which
give the approximate relation $P_{\Omega}\simeq(5/3)P_{\Lambda}$
for primary hyperons and include feed-down effects for
$\Lambda$~\cite{Li:2021zwq}.
The difference between the measurements and the model prediction could potentially arise from quark--quark spin correlations contributing to the $\Omega$ polarization~\cite{Zhang:2024hyq,Oliva:2026wbo}.
These correlations are sensitive to the hyperon production mechanisms, the dynamical evolution of the system, and potential spin--momentum or spin--vorticity interferences. 
Furthermore, the difference between the $\Lambda$ and $\Omega$ polarizations could also arise from different production times or regions as the system freezes out~\cite{Vitiuk:2019rfv}, meson--baryon interactions~\cite{Csernai:2018yok}, 
or the magnetic--field contribution to the polarization due to the larger magnetic moment of the $\Omega$ hyperon.


In summary, we report measurements of the global polarization of $\Lambda$, $\bar{\Lambda}$, $\Xi^{-}+\bar{\Xi}^{+}$, and $\Omega^{-}+\bar{\Omega}^{+}$ hyperons in Au+Au collisions at $\sqrt{s_{\mathrm{NN}}}=7.7$, 9.2, 11.5, 14.6, 17.3, 19.6, and 27~GeV with high-statistics BES-II data at STAR. 
No significant difference is observed between $\Lambda$ and $\bar{\Lambda}$ polarization across a wider collision energy range. For the first time, evidence for a positive rapidity dependence is observed for $\Lambda$ global polarization at lower collision energies, providing new constraints on the longitudinal vorticity structure and the interplay between thermal vorticity and shear flow.
For the first time, significant polarization of $\Xi^{-}+\bar{\Xi}^{+}$ is observed at lower collision energies, following a decreasing trend versus energy similar to that of $\Lambda+\bar{\Lambda}$. 
The $\Xi^{-}+\bar{\Xi}^{+}$ polarizations measured via the daughter-$\Lambda$ angular distribution and via spin transfer to the daughter-$\Lambda$ are consistent and comparable in magnitude to those of $\Lambda+\bar{\Lambda}$.
There is an indication that $\Omega$ polarization also decreases with collision energy as inferred from the indirect measurement under the assumption $\gamma_\Omega=1$.
A hint of larger $\Omega$ polarization than that of inclusive $\Lambda$ is observed at lower energies with relatively large uncertainties.
Such a unified energy dependence trend for $\Lambda$, $\Xi$, and $\Omega$ supports a common vorticity-driven origin of global polarization.
Further theoretical developments will help clarify the species and rapidity dependence of global polarization, while higher-precision measurements would provide additional tests.
The present measurements provide a coherent set of global-polarization observables spanning collision energy, hyperon species, rapidity, and particle--antiparticle differences.


\begin{acknowledgments}
We thank the RHIC Operations Group and SCDF at BNL, the NERSC Center at LBNL, and the Open Science Grid consortium for providing resources and support.  This work was supported in part by the Office of Nuclear Physics within the U.S. DOE Office of Science, the U.S. National Science Foundation, National Natural Science Foundation of China, Chinese Academy of Science, the Ministry of Science and Technology of China and the Chinese Ministry of Education, NSTC Taipei, the National Research Foundation of Korea, Czech Science Foundation and Ministry of Education, Youth and Sports of the Czech Republic, Hungarian National Research, Development and Innovation Office, New National Excellency Programme of the Hungarian Ministry of Human Capacities, Department of Atomic Energy and Department of Science and Technology of the Government of India, the National Science Centre and WUT ID-UB of Poland, German Bundesministerium f\"ur Bildung, Wissenschaft, Forschung and Technologie (BMBF), Helmholtz Association, Ministry of Education, Culture, Sports, Science, and Technology (MEXT), Japan Society for the Promotion of Science (JSPS), and Agencia Nacional de Investigacion y Desarrollo de Chile (ANID), Chile.
\end{acknowledgments}

\bibliography{refs}

@article{1,
    author = "Liang, Zuo-Tang and Wang, Xin-Nian",
    title = "{Globally polarized quark-gluon plasma in non-central A+A collisions}",
    eprint = "nucl-th/0410079",
    archivePrefix = "arXiv",
    reportNumber = "LBNL-56383",
    doi = "10.1103/PhysRevLett.94.102301",
    journal = "Phys. Rev. Lett.",
    volume = "94",
    pages = "102301",
    year = "2005",
    note = "[Erratum: Phys.Rev.Lett. 96, 039901 (2006)]"
}

@article{2,
    author = "Becattini, F. and Piccinini, F. and Rizzo, J.",
    title = "{Angular momentum conservation in heavy ion collisions at very high energy}",
    eprint = "0711.1253",
    archivePrefix = "arXiv",
    primaryClass = "nucl-th",
    doi = "10.1103/PhysRevC.77.024906",
    journal = "Phys. Rev. C",
    volume = "77",
    pages = "024906",
    year = "2008"
}

@article{STAR:2017ckg,
    author = "Adamczyk, L. and others",
    collaboration = "STAR",
    title = "{Global $\Lambda$ hyperon polarization in nuclear collisions: evidence for the most vortical fluid}",
    eprint = "1701.06657",
    archivePrefix = "arXiv",
    primaryClass = "nucl-ex",
    doi = "10.1038/nature23004",
    journal = "Nature",
    volume = "548",
    pages = "62--65",
    year = "2017"
}

@article{STAR:2022fan,
    author = "Abdallah, M. S. and others",
    collaboration = "STAR",
    title = "{Pattern of global spin alignment of \ensuremath{\phi} and K$^{*0}$ mesons in heavy-ion collisions}",
    eprint = "2204.02302",
    archivePrefix = "arXiv",
    primaryClass = "hep-ph",
    doi = "10.1038/s41586-022-05557-5",
    journal = "Nature",
    volume = "614",
    number = "7947",
    pages = "244--248",
    year = "2023"
}

@article{Chen:2026gka,
    author = "Chen, Jinhui and others",
    title = "{Selected Highlights from STAR Experiment}",
    eprint = "2601.12977",
    archivePrefix = "arXiv",
    primaryClass = "nucl-ex",
    doi = "10.1088/0256-307X/43/3/030102",
    journal = "Chin. Phys. Lett.",
    volume = "43",
    number = "3",
    pages = "030102",
    year = "2026"
}

@article{STAR:2005gfr,
    author = "Adams, John and others",
    collaboration = "STAR",
    title = "{Experimental and theoretical challenges in the search for the quark gluon plasma: The STAR Collaboration's critical assessment of the evidence from RHIC collisions}",
    eprint = "nucl-ex/0501009",
    archivePrefix = "arXiv",
    doi = "10.1016/j.nuclphysa.2005.03.085",
    journal = "Nucl. Phys. A",
    volume = "757",
    pages = "102--183",
    year = "2005"
}

@article{PHENIX:2004vcz,
    author = "Adcox, K. and others",
    collaboration = "PHENIX",
    title = "{Formation of dense partonic matter in relativistic nucleus-nucleus collisions at RHIC: Experimental evaluation by the PHENIX collaboration}",
    eprint = "nucl-ex/0410003",
    archivePrefix = "arXiv",
    doi = "10.1016/j.nuclphysa.2005.03.086",
    journal = "Nucl. Phys. A",
    volume = "757",
    pages = "184--283",
    year = "2005"
}

@article{BRAHMS:2004adc,
    author = "Arsene, I. and others",
    collaboration = "BRAHMS",
    title = "{Quark gluon plasma and color glass condensate at RHIC? The Perspective from the BRAHMS experiment}",
    eprint = "nucl-ex/0410020",
    archivePrefix = "arXiv",
    doi = "10.1016/j.nuclphysa.2005.02.130",
    journal = "Nucl. Phys. A",
    volume = "757",
    pages = "1--27",
    year = "2005"
}

@article{PHOBOS:2004zne,
    author = "Back, B. B. and others",
    collaboration = "PHOBOS",
    title = "{The PHOBOS perspective on discoveries at RHIC}",
    eprint = "nucl-ex/0410022",
    archivePrefix = "arXiv",
    doi = "10.1016/j.nuclphysa.2005.03.084",
    journal = "Nucl. Phys. A",
    volume = "757",
    pages = "28--101",
    year = "2005"
}

@article{Heinz:2013th,
    author = "Heinz, Ulrich and Snellings, Raimond",
    title = "{Collective flow and viscosity in relativistic heavy-ion collisions}",
    eprint = "1301.2826",
    archivePrefix = "arXiv",
    primaryClass = "nucl-th",
    doi = "10.1146/annurev-nucl-102212-170540",
    journal = "Ann. Rev. Nucl. Part. Sci.",
    volume = "63",
    pages = "123--151",
    year = "2013"
}

@article{Bernhard:2019bmu,
    author = "Bernhard, Jonah E. and Moreland, J. Scott and Bass, Steffen A.",
    title = "{Bayesian estimation of the specific shear and bulk viscosity of quark{\textendash}gluon plasma}",
    doi = "10.1038/s41567-019-0611-8",
    journal = "Nature Phys.",
    volume = "15",
    number = "11",
    pages = "1113--1117",
    year = "2019"
}

@article{ALICE:2022wpn,
    author = "Acharya, Shreyasi and others",
    collaboration = "ALICE",
    title = "{The ALICE experiment: a journey through QCD}",
    eprint = "2211.04384",
    archivePrefix = "arXiv",
    primaryClass = "nucl-ex",
    reportNumber = "CERN-EP-2022-227",
    doi = "10.1140/epjc/s10052-024-12935-y",
    journal = "Eur. Phys. J. C",
    volume = "84",
    number = "8",
    pages = "813",
    year = "2024"
}

@article{STAR:2020xbm,
    author = "Adam, J. and others",
    collaboration = "STAR",
    title = "{Global Polarization of $\Xi$ and $\Omega$ Hyperons in Au+Au Collisions at $\sqrt {s_{NN}}$ = 200  GeV}",
    eprint = "2012.13601",
    archivePrefix = "arXiv",
    primaryClass = "nucl-ex",
    doi = "10.1103/PhysRevLett.126.162301",
    journal = "Phys. Rev. Lett.",
    volume = "126",
    number = "16",
    pages = "162301",
    year = "2021"
}

@article{STAR:2007ccu,
    author = "Abelev, B. I. and others",
    collaboration = "STAR",
    title = "{Global polarization measurement in Au+Au collisions}",
    eprint = "0705.1691",
    archivePrefix = "arXiv",
    primaryClass = "nucl-ex",
    reportNumber = "STAR-05-11-2007",
    doi = "10.1103/PhysRevC.76.024915",
    journal = "Phys. Rev. C",
    volume = "76",
    pages = "024915",
    year = "2007",
    note = "[Erratum: Phys.Rev.C 95, 039906 (2017)]"
}

@article{STAR:2018gyt,
    author = "Adam, Jaroslav and others",
    collaboration = "STAR",
    title = "{Global polarization of $\Lambda$ hyperons in Au+Au collisions at $\sqrt{s_{_{NN}}}$ = 200 GeV}",
    eprint = "1805.04400",
    archivePrefix = "arXiv",
    primaryClass = "nucl-ex",
    doi = "10.1103/PhysRevC.98.014910",
    journal = "Phys. Rev. C",
    volume = "98",
    pages = "014910",
    year = "2018"
}

@article{HADES:2022enx,
    author = "Abou Yassine, R. and others",
    collaboration = "HADES",
    title = "{Measurement of global polarization of $\Lambda$ hyperons in few-GeV heavy-ion collisions}",
    eprint = "2207.05160",
    archivePrefix = "arXiv",
    primaryClass = "nucl-ex",
    doi = "10.1016/j.physletb.2022.137506",
    journal = "Phys. Lett. B",
    volume = "835",
    pages = "137506",
    year = "2022"
}

@article{ALICE:2019onw,
    author = "Acharya, Shreyasi and others",
    collaboration = "ALICE",
    title = "{Global polarization of $\Lambda \bar \Lambda$ hyperons in Pb-Pb collisions at $\sqrt {s_{NN}}$ = 2.76 and 5.02 TeV}",
    eprint = "1909.01281",
    archivePrefix = "arXiv",
    primaryClass = "nucl-ex",
    reportNumber = "CERN-EP-2019-173",
    doi = "10.1103/PhysRevC.101.044611",
    journal = "Phys. Rev. C",
    volume = "101",
    number = "4",
    pages = "044611",
    year = "2020",
    note = "[Erratum: Phys.Rev.C 105, 029902 (2022)]"
}

@article{STAR:2021beb,
    author = "Abdallah, M. S. and others",
    collaboration = "STAR",
    title = "{Global $\Lambda$-hyperon polarization in Au+Au collisions at $\sqrt {s_{NN}}$=3~GeV}",
    eprint = "2108.00044",
    archivePrefix = "arXiv",
    primaryClass = "nucl-ex",
    doi = "10.1103/PhysRevC.104.L061901",
    journal = "Phys. Rev. C",
    volume = "104",
    number = "6",
    pages = "L061901",
    year = "2021"
}

@article{STAR:2002eio,
    author = "Ackermann, K. H. and others",
    collaboration = "STAR",
    title = "{STAR detector overview}",
    doi = "10.1016/S0168-9002(02)01960-5",
    journal = "Nucl. Instrum. Meth. A",
    volume = "499",
    pages = "624--632",
    year = "2003"
}

@article{Wu:2022mkr,
    author = "Wu, Xiang-Yu and Yi, Cong and Qin, Guang-You and Pu, Shi",
    title = "{Local and global polarization of {$\Lambda$} hyperons across RHIC-BES energies: The roles of spin hall effect, initial condition, and baryon diffusion}",
    eprint = "2204.02218",
    archivePrefix = "arXiv",
    primaryClass = "hep-ph",
    doi = "10.1103/PhysRevC.105.064909",
    journal = "Phys. Rev. C",
    volume = "105",
    number = "6",
    pages = "064909",
    year = "2022"
}

@misc{Li:2026udm,
    author = "Li, Gen-Hui and Yi, Cong and Wu, Xiang-Yu and Pu, Shi and Qin, Guang-You",
    title = "{Global polarization of $\Lambda$, $\Xi^{-}$, and $\Omega^{-}$ hyperons in Au+Au collisions at RHIC BES-II energies}",
    eprint = "2606.10341",
    archivePrefix = "arXiv",
    primaryClass = "nucl-th",
    month = "6",
    year = "2026"
}

@article{Anderson:2003ur,
    author = "Anderson, M. and others",
    title = "{The Star time projection chamber: A Unique tool for studying high multiplicity events at RHIC}",
    eprint = "nucl-ex/0301015",
    archivePrefix = "arXiv",
    doi = "10.1016/S0168-9002(02)01964-2",
    journal = "Nucl. Instrum. Meth. A",
    volume = "499",
    pages = "659--678",
    year = "2003"
}

@article{Gao:2007bc,
    author = "Gao, Jian-Hua and Chen, Shou-Wan and Deng, Wei-Tian and Liang, Zuo-Tang and Wang, Qun and Wang, Xin-Nian",
    title = "{Global quark polarization in non-central A+A collisions}",
    eprint = "0710.2943",
    archivePrefix = "arXiv",
    primaryClass = "nucl-th",
    reportNumber = "LBNL-63515",
    doi = "10.1103/PhysRevC.77.044902",
    journal = "Phys. Rev. C",
    volume = "77",
    pages = "044902",
    year = "2008"
}

@article{Chen:2024aom,
    author = "Chen, Jinhui and others",
    title = "{Properties of the QCD matter: review of selected results from the relativistic heavy ion collider beam energy scan (RHIC BES) program}",
    eprint = "2407.02935",
    archivePrefix = "arXiv",
    primaryClass = "nucl-ex",
    doi = "10.1007/s41365-024-01591-2",
    journal = "Nucl. Sci. Tech.",
    volume = "35",
    number = "12",
    pages = "214",
    year = "2024"
}

@article{Llope:2014nva,
    author = "Llope, W. J. and others",
    title = "{The STAR Vertex Position Detector}",
    eprint = "1403.6855",
    archivePrefix = "arXiv",
    primaryClass = "physics.ins-det",
    doi = "10.1016/j.nima.2014.04.080",
    journal = "Nucl. Instrum. Meth. A",
    volume = "759",
    pages = "23--28",
    year = "2014"
}

@article{Adler:2001fq,
    author = "Adler, C. and Strobele, H. and Denisov, A. and Garcia, E. and Murray, M. and White, S.",
    editor = "Batignani, G. and Cervelli, F. and Chiarelli, G. and Scribano, A.",
    title = "{The RHIC zero-degree calorimeters}",
    doi = "10.1016/S0168-9002(00)01238-9",
    journal = "Nucl. Instrum. Meth. A",
    volume = "461",
    pages = "337--340",
    year = "2001"
}

@article{Voloshin:2008dg,
    author = "Voloshin, Sergei A. and Poskanzer, Arthur M. and Snellings, Raimond",
    editor = "Stock, R.",
    title = "{Collective phenomena in non-central nuclear collisions}",
    eprint = "0809.2949",
    archivePrefix = "arXiv",
    primaryClass = "nucl-ex",
    doi = "10.1007/978-3-642-01539-7_10",
    journal = "Landolt-Bornstein",
    volume = "23",
    pages = "293--333",
    year = "2010"
}

@article{Adams:2019fpo,
    author = "Adams, Joseph and others",
    title = "{The STAR Event Plane Detector}",
    eprint = "1912.05243",
    archivePrefix = "arXiv",
    primaryClass = "physics.ins-det",
    doi = "10.1016/j.nima.2020.163970",
    journal = "Nucl. Instrum. Meth. A",
    volume = "968",
    pages = "163970",
    year = "2020"
}

@article{Adams:2021yob,
    author = "Adams, Joseph R. and Lisa, Michael A.",
    title = "{Decorrelation of participant and spectator angular momenta in heavy-ion collisions}",
    eprint = "2109.14726",
    archivePrefix = "arXiv",
    primaryClass = "nucl-th",
    doi = "10.1103/PhysRevC.106.064904",
    journal = "Phys. Rev. C",
    volume = "106",
    number = "6",
    pages = "064904",
    year = "2022"
}

@article{Llope:2012zz,
    author = "Llope, W. J.",
    editor = "Herrmann, Norbert and Santonico, Rinaldo and Ratti, Sergio P.",
    collaboration = "STAR",
    title = "{Multigap RPCs in the STAR experiment at RHIC}",
    doi = "10.1016/j.nima.2010.07.086",
    journal = "Nucl. Instrum. Meth. A",
    volume = "661",
    pages = "S110--S113",
    year = "2012"
}

@article{Poskanzer:1998yz,
    author = "Poskanzer, Arthur M. and Voloshin, S. A.",
    title = "{Methods for analyzing anisotropic flow in relativistic nuclear collisions}",
    eprint = "nucl-ex/9805001",
    archivePrefix = "arXiv",
    doi = "10.1103/PhysRevC.58.1671",
    journal = "Phys. Rev. C",
    volume = "58",
    pages = "1671--1678",
    year = "1998"
}

@article{Deng:2016gyh,
    author = "Deng, Wei-Tian and Huang, Xu-Guang",
    title = "{Vorticity in Heavy-Ion Collisions}",
    eprint = "1603.06117",
    archivePrefix = "arXiv",
    primaryClass = "nucl-th",
    doi = "10.1103/PhysRevC.93.064907",
    journal = "Phys. Rev. C",
    volume = "93",
    number = "6",
    pages = "064907",
    year = "2016"
}

@article{Wei:2018zfb,
    author = "Wei, De-Xian and Deng, Wei-Tian and Huang, Xu-Guang",
    title = "{Thermal vorticity and spin polarization in heavy-ion collisions}",
    eprint = "1810.00151",
    archivePrefix = "arXiv",
    primaryClass = "nucl-th",
    doi = "10.1103/PhysRevC.99.014905",
    journal = "Phys. Rev. C",
    volume = "99",
    number = "1",
    pages = "014905",
    year = "2019"
}

@article{Guo:2021udq,
    author = "Guo, Yu and Liao, Jinfeng and Wang, Enke and Xing, Hongxi and Zhang, Hui",
    title = "{Hyperon polarization from the vortical fluid in low-energy nuclear collisions}",
    eprint = "2105.13481",
    archivePrefix = "arXiv",
    primaryClass = "nucl-th",
    doi = "10.1103/PhysRevC.104.L041902",
    journal = "Phys. Rev. C",
    volume = "104",
    number = "4",
    pages = "L041902",
    year = "2021"
}

@article{Wu:2019eyi,
    author = "Wu, Hong-Zhong and Pang, Long-Gang and Huang, Xu-Guang and Wang, Qun",
    title = "{Local spin polarization in high energy heavy ion collisions}",
    eprint = "1906.09385",
    archivePrefix = "arXiv",
    primaryClass = "nucl-th",
    doi = "10.1103/PhysRevResearch.1.033058",
    journal = "Phys. Rev. Research.",
    volume = "1",
    pages = "033058",
    year = "2019"
}

@article{Kharzeev:2007jp,
    author = "Kharzeev, Dmitri E. and McLerran, Larry D. and Warringa, Harmen J.",
    title = "{The Effects of topological charge change in heavy ion collisions: 'Event by event P and CP violation'}",
    eprint = "0711.0950",
    archivePrefix = "arXiv",
    primaryClass = "hep-ph",
    doi = "10.1016/j.nuclphysa.2008.02.298",
    journal = "Nucl. Phys. A",
    volume = "803",
    pages = "227--253",
    year = "2008"
}

@article{Skokov:2009qp,
    author = "Skokov, V. and Illarionov, A. Yu. and Toneev, V.",
    title = "{Estimate of the magnetic field strength in heavy-ion collisions}",
    eprint = "0907.1396",
    archivePrefix = "arXiv",
    primaryClass = "nucl-th",
    doi = "10.1142/S0217751X09047570",
    journal = "Int. J. Mod. Phys. A",
    volume = "24",
    pages = "5925--5932",
    year = "2009"
}

@article{Voronyuk:2011jd,
    author = "Voronyuk, V. and Toneev, V. D. and Cassing, W. and Bratkovskaya, E. L. and Konchakovski, V. P. and Voloshin, S. A.",
    title = "{(Electro-)Magnetic field evolution in relativistic heavy-ion collisions}",
    eprint = "1103.4239",
    archivePrefix = "arXiv",
    primaryClass = "nucl-th",
    doi = "10.1103/PhysRevC.83.054911",
    journal = "Phys. Rev. C",
    volume = "83",
    pages = "054911",
    year = "2011"
}

@article{Liang:2019pst,
    author = "Liang, Zuo-Tang and Song, Jun and Upsal, Isaac and Wang, Qun and Xu, Zhang-Bu",
    title = "{Rapidity dependence of global polarization in heavy ion collisions}",
    eprint = "1912.10223",
    archivePrefix = "arXiv",
    primaryClass = "nucl-th",
    doi = "10.1088/1674-1137/abc065",
    journal = "Chin. Phys. C",
    volume = "45",
    number = "1",
    pages = "014102",
    year = "2021"
}

@article{Jiang:2016woz,
    author = "Jiang, Yin and Lin, Zi-Wei and Liao, Jinfeng",
    title = "{Rotating quark-gluon plasma in relativistic heavy ion collisions}",
    eprint = "1602.06580",
    archivePrefix = "arXiv",
    primaryClass = "hep-ph",
    doi = "10.1103/PhysRevC.94.044910",
    journal = "Phys. Rev. C",
    volume = "94",
    number = "4",
    pages = "044910",
    year = "2016",
    note = "[Erratum: Phys.Rev.C 95, 049904 (2017)]"
}

@article{Becattini:2016gvu,
    author = "Becattini, F. and Karpenko, I. and Lisa, M. and Upsal, I. and Voloshin, S.",
    title = "{Global hyperon polarization at local thermodynamic equilibrium with vorticity, magnetic field and feed-down}",
    eprint = "1610.02506",
    archivePrefix = "arXiv",
    primaryClass = "nucl-th",
    doi = "10.1103/PhysRevC.95.054902",
    journal = "Phys. Rev. C",
    volume = "95",
    number = "5",
    pages = "054902",
    year = "2017"
}

@article{Muller:2018ibh,
    author = {M\"uller, Berndt and Sch\"afer, Andreas},
    title = "{Chiral magnetic effect and an experimental bound on the late time magnetic field strength}",
    eprint = "1806.10907",
    archivePrefix = "arXiv",
    primaryClass = "hep-ph",
    doi = "10.1103/PhysRevD.98.071902",
    journal = "Phys. Rev. D",
    volume = "98",
    number = "7",
    pages = "071902",
    year = "2018"
}

@article{STAR:2023nvo,
    author = "Abdulhamid, M. I. and others",
    collaboration = "STAR",
    title = "{Global polarization of $\Lambda$ and $\bar\Lambda$ hyperons in Au+Au collisions at $\sqrt{s_{\mathrm{NN}}}$=19.6 and 27 GeV}",
    eprint = "2305.08705",
    archivePrefix = "arXiv",
    primaryClass = "nucl-ex",
    doi = "10.1103/PhysRevC.108.014910",
    journal = "Phys. Rev. C",
    volume = "108",
    number = "1",
    pages = "014910",
    year = "2023"
}

@article{Vitiuk:2019rfv,
    author = "Vitiuk, O. and Bravina, L. V. and Zabrodin, E. E.",
    title = "{Is different $\Lambda$ and $\bar \Lambda$ polarization caused by different spatio-temporal freeze-out picture?}",
    eprint = "1910.06292",
    archivePrefix = "arXiv",
    primaryClass = "hep-ph",
    doi = "10.1016/j.physletb.2020.135298",
    journal = "Phys. Lett. B",
    volume = "803",
    pages = "135298",
    year = "2020"
}

@article{Csernai:2018yok,
    author = "Csernai, L. P. and Kapusta, J. I. and Welle, T.",
    title = "{$\Lambda$ and $\bar{\Lambda}$ spin interaction with meson fields generated by the baryon current in high energy nuclear collisions}",
    eprint = "1807.11521",
    archivePrefix = "arXiv",
    primaryClass = "nucl-th",
    doi = "10.1103/PhysRevC.99.021901",
    journal = "Phys. Rev. C",
    volume = "99",
    number = "2",
    pages = "021901",
    year = "2019"
}

@article{Xie:2019jun,
    author = "Xie, Yilong and Wang, Dujuan and Csernai, Laszlo Pal",
    title = "{Fluid dynamics study of the $\varLambda $ polarization for Au + Au collisions at $\sqrt{s_{NN}}=200$ GeV}",
    eprint = "1907.00773",
    archivePrefix = "arXiv",
    primaryClass = "hep-ph",
    doi = "10.1140/epjc/s10052-019-7576-8",
    journal = "Eur. Phys. J. C",
    volume = "80",
    number = "1",
    pages = "39",
    year = "2020"
}

@article{Li:2021zwq,
    author = "Li, Hui and Xia, Xiao-Liang and Huang, Xu-Guang and Huang, Huan Zhong",
    title = "{Global spin polarization of multistrange hyperons and feed-down effect in heavy-ion collisions}",
    eprint = "2106.09443",
    archivePrefix = "arXiv",
    primaryClass = "nucl-th",
    doi = "10.1016/j.physletb.2022.136971",
    journal = "Phys. Lett. B",
    volume = "827",
    pages = "136971",
    year = "2022"
}

@article{Becattini:2020ngo,
    author = "Becattini, Francesco and Lisa, Michael A.",
    title = "{Polarization and Vorticity in the Quark\textendash{}Gluon Plasma}",
    eprint = "2003.03640",
    archivePrefix = "arXiv",
    primaryClass = "nucl-ex",
    doi = "10.1146/annurev-nucl-021920-095245",
    journal = "Ann. Rev. Nucl. Part. Sci.",
    volume = "70",
    pages = "395--423",
    year = "2020"
}

@article{Fang:2023bbw,
    author = "Fang, Shuo and Pu, Shi and Yang, Di-Lun",
    title = "{Spin polarization and spin alignment from quantum kinetic theory with self-energy corrections}",
    eprint = "2311.15197",
    archivePrefix = "arXiv",
    primaryClass = "hep-ph",
    doi = "10.1103/PhysRevD.109.034034",
    journal = "Phys. Rev. D",
    volume = "109",
    number = "3",
    pages = "034034",
    year = "2024"
}

@article{Karpenko:2016jyx,
    author = "Karpenko, I. and Becattini, F.",
    title = "{Study of $\Lambda $ polarization in relativistic nuclear collisions at $\sqrt{s_\mathrm {NN}}=7.7$ \textendash{}200 GeV}",
    eprint = "1610.04717",
    archivePrefix = "arXiv",
    primaryClass = "nucl-th",
    doi = "10.1140/epjc/s10052-017-4765-1",
    journal = "Eur. Phys. J. C",
    volume = "77",
    number = "4",
    pages = "213",
    year = "2017"
}

@article{Niida:2024ntm,
    author = "Niida, Takafumi and Voloshin, Sergei A.",
    title = "{Polarization phenomenon in heavy-ion collisions}",
    eprint = "2404.11042",
    archivePrefix = "arXiv",
    primaryClass = "nucl-ex",
    doi = "10.1142/S0218301324300108",
    journal = "Int. J. Mod. Phys. E",
    volume = "33",
    number = "09",
    pages = "2430010",
    year = "2024"
}

@article{Zhang:2024hyq,
    author = "Zhang, Zhe and Lv, Ji-peng and Yu, Zi-han and Liang, Zuo-tang",
    title = "{Global tensor polarization of spin 3/2 hadrons and quark spin correlations in relativistic heavy ion collisions}",
    eprint = "2406.03840",
    archivePrefix = "arXiv",
    primaryClass = "hep-ph",
    doi = "10.1103/PhysRevD.110.074019",
    journal = "Phys. Rev. D",
    volume = "110",
    number = "7",
    pages = "074019",
    year = "2024"
}

@article{Guo:2019joy,
    author = "Guo, Yu and Shi, Shuzhe and Feng, Shengqin and Liao, Jinfeng",
    title = "{Magnetic Field Induced Polarization Difference between Hyperons and Anti-hyperons}",
    eprint = "1905.12613",
    archivePrefix = "arXiv",
    primaryClass = "nucl-th",
    doi = "10.1016/j.physletb.2019.134929",
    journal = "Phys. Lett. B",
    volume = "798",
    pages = "134929",
    year = "2019"
}

@article{STAR:2012och,
    author = "Adamczyk, L. and others",
    collaboration = "STAR",
    title = "{Inclusive charged hadron elliptic flow in Au + Au collisions at $\sqrt{s_{NN}}$ = 7.7 - 39 GeV}",
    eprint = "1206.5528",
    archivePrefix = "arXiv",
    primaryClass = "nucl-ex",
    doi = "10.1103/PhysRevC.86.054908",
    journal = "Phys. Rev. C",
    volume = "86",
    pages = "054908",
    year = "2012"
}

@article{Baznat:2016rapidity,
  title = {Femto-vortex sheets and hyperon polarization in heavy-ion collisions},
  author = {Baznat, Mircea and Gudima, Konstantin and Sorin, Alexander and Teryaev, Oleg},
  journal = {Phys. Rev. C},
  volume = {93},
  issue = {3},
  pages = {031902},
  numpages = {5},
  year = {2016},
  month = {Mar},
  publisher = {American Physical Society},
  doi = {10.1103/PhysRevC.93.031902},
  url = {https://link.aps.org/doi/10.1103/PhysRevC.93.031902}
}

@article{Sheng:2024pbw,
    author = "Sheng, Xin-Li and Becattini, Francesco and Huang, Xu-Guang and Zhang, Zhong-Hua",
    title = "{Spin polarization of fermions at local equilibrium: Second-order gradient expansion}",
    eprint = "2407.12130",
    archivePrefix = "arXiv",
    primaryClass = "hep-th",
    doi = "10.1103/PhysRevC.110.064908",
    journal = "Phys. Rev. C",
    volume = "110",
    number = "6",
    pages = "064908",
    year = "2024"
}

@article{Becattini:2024uha,
    author = "Becattini, Francesco and Buzzegoli, Matteo and Niida, Takafumi and Pu, Shi and Tang, Ai-Hong and Wang, Qun",
    title = "{Spin polarization in relativistic heavy-ion collisions}",
    eprint = "2402.04540",
    archivePrefix = "arXiv",
    primaryClass = "nucl-th",
    doi = "10.1142/9789811294679_0005",
    journal = "Int. J. Mod. Phys. E",
    volume = "33",
    number = "06",
    pages = "2430006",
    year = "2024"
}

@article{Yi:2024kwu,
    author = "Yi, Cong and Wu, Xiang-Yu and Zhu, Jie and Pu, Shi and Qin, Guang-You",
    title = "{Spin polarization of {$\Lambda$} hyperons along the beam direction in p+Pb collisions at $\sqrt{s_{\mathrm{NN}}}$=8.16 TeV using hydrodynamic approaches}",
    eprint = "2408.04296",
    archivePrefix = "arXiv",
    primaryClass = "hep-ph",
    doi = "10.1103/PhysRevC.111.044901",
    journal = "Phys. Rev. C",
    volume = "111",
    number = "4",
    pages = "044901",
    year = "2025"
}

@article{Sun:2017xhx,
    author = "Sun, Yifeng and Ko, Che Ming",
    title = "{$\Lambda$ hyperon polarization in relativistic heavy ion collisions from a chiral kinetic approach}",
    eprint = "1706.09467",
    archivePrefix = "arXiv",
    primaryClass = "nucl-th",
    doi = "10.1103/PhysRevC.96.024906",
    journal = "Phys. Rev. C",
    volume = "96",
    number = "2",
    pages = "024906",
    year = "2017"
}

@article{Voronyuk:2023vyu,
    author = "Voronyuk, V. and Tsegelnik, N. S. and Kolomeitsev, E. E.",
    title = "{Hyperon global polarization in heavy-ion collisions at energies available at the JINR Nuclotron-based Ion Collider fAcility: Feed-down effects and the role of {\ensuremath{\Sigma}}0 hyperons}",
    eprint = "2305.10792",
    archivePrefix = "arXiv",
    primaryClass = "nucl-th",
    doi = "10.1103/PhysRevC.111.034907",
    journal = "Phys. Rev. C",
    volume = "111",
    number = "3",
    pages = "034907",
    year = "2025"
}

@article{HyperCP:2004not,
    author = "Huang, M. and others",
    collaboration = "HyperCP",
    title = "{New measurement of $\Xi^{-}\to\Lambda\pi^{-}$ decay parameters}",
    reportNumber = "FERMILAB-PUB-04-347-E",
    doi = "10.1103/PhysRevLett.93.011802",
    journal = "Phys. Rev. Lett.",
    volume = "93",
    pages = "011802",
    year = "2004"
}

@article{E756:2000rge,
    author = "Luk, K. B. and others",
    collaboration = "E756",
    title = "{Search for direct CP violation in nonleptonic decays of charged Xi and Lambda hyperons}",
    eprint = "hep-ex/0007030",
    archivePrefix = "arXiv",
    reportNumber = "LBNL-46257, LBL-46257, FERMILAB-PUB-00-410-A",
    doi = "10.1103/PhysRevLett.85.4860",
    journal = "Phys. Rev. Lett.",
    volume = "85",
    pages = "4860--4863",
    year = "2000"
}

@article{STAR:2025dgs,
    author = "Aboona, B. E. and others",
    collaboration = "STAR",
    title = "{Hyperon global polarization in isobar Ru+Ru and Zr+Zr collisions at $\sqrt{s_{\mathrm{NN}}}$=200GeV}",
    eprint = "2505.05046",
    archivePrefix = "arXiv",
    primaryClass = "nucl-ex",
    doi = "10.1016/j.physletb.2025.139891",
    journal = "Phys. Lett. B",
    volume = "870",
    pages = "139891",
    year = "2025"
}

@article{Shen:2018pkc,
    author = "Shen, Fuwang and Wang, Shuai and Kong, Fangang and Bai, Shiwei and Li, Changyu and Videb{\ae}k, Flemming and Xu, Zhangbu and Zhu, Chengguang and Xu, Qinghua and Yang, Chi",
    title = "{MWPC prototyping and performance test for the STAR inner TPC upgrade}",
    eprint = "1805.03938",
    archivePrefix = "arXiv",
    primaryClass = "physics.ins-det",
    doi = "10.1016/j.nima.2018.04.019",
    journal = "Nucl. Instrum. Meth. A",
    volume = "896",
    pages = "90--95",
    year = "2018"
}

@unpublished{longpaper,
  author = "{STAR Collaboration}",
  title = "{Differential measurement of global polarization of $\Lambda$, $\Xi$, and $\Omega$ hyperons in Au+Au collisions from the RHIC Beam Energy Scan-II}",
  eprint        = {arXiv:XXXX.XXXXX},
  archivePrefix = {arXiv},
  primaryClass  = {nucl-ex},
  note = "{Companion paper XXX}",
  year = {2026}
}

@article{Oliva:2026wbo,
    author = "Oliva, Lucia and Wang, Qun and Wang, Xin-Nian",
    title = "{Quark spin correlation inside hyperons}",
    eprint = "2603.10427",
    archivePrefix = "arXiv",
    primaryClass = "hep-ph",
    doi = "10.1103/h88t-phwj",
    journal = "Phys. Rev. C",
    volume = "114",
    number = "1",
    pages = "014914",
    year = "2026"
}

@article{ParticleDataGroup:2020ssz,
    author = "Zyla, P. A. and others",
    collaboration = "Particle Data Group",
    title = "{Review of Particle Physics}",
    doi = "10.1093/ptep/ptaa104",
    journal = "PTEP",
    volume = "2020",
    number = "8",
    pages = "083C01",
    year = "2020"
}

\end{document}